\documentstyle[11pt]{article}
\def\bkR{{\rm I\kern-.17em R}}
\def\bkC{{\rm \kern.24em \vrule width.05em height1.4ex depth-.05ex \kern-.26em C}}

\def\bkR{{\rm I\kern-.17em R}}
\def\bkC{{\rm \kern.24em \vrule width.05em height1.4ex depth-.05ex \kern-.26em C}}

\begin{document}

\author{Nuno Costa Dias\footnote{{\it ncdias@meo.pt}} \\ Jo\~{a}o Nuno Prata\footnote{{\it joao.prata@mail.telepac.pt}} \\ {\it Departamento de Matem\'atica} \\
{\it Universidade Lus\'ofona de Humanidades e Tecnologias} \\ {\it Av. Campo Grande, 376, 1749-024 Lisboa, Portugal}\\
{\it and}\\
{\it Grupo de F\'{\i}sica Matem\'atica}\\
{\it Universidade de Lisboa}\\
{\it Av. Prof. Gama Pinto 2}\\
{\it 1649-003 Lisboa, Potugal}}

\title{The Narcowich-Wigner spectrum of a pure state}

\maketitle
\begin{abstract}
We completely characterize the Narcowich-Wigner spectrum of Wigner functions associated with pure states.
\end{abstract}

\section{Introduction}

In the context of the Weyl-Wigner formulation of quantum mechanics, the derivation of criteria characterizing the phase space functions in terms of their quantum mechanical properties is a long standing problem. Such criteria should, at least, provide a simple and practical
answer to the following questions: i) Which square integrable phase space functions are Wigner functions?; ii) Which Wigner functions are
associated with pure states?; iii) Which Wigner functions are associated with mixed states? and iv) Which Wigner functions are non-negative?

Partial answers to these questions have been provided in the literature: Hudson \cite{Hudson} proved that Wigner functions of pure one
dimensional states are everywhere non-negative iff the state is coherent. Soto and Claverie \cite{Soto} generalized this result for higher
dimensional systems. However, the analogous characterization of positive Wigner functions associated with mixed states remains an open question. Some results in this direction were obtained in \cite{Simon} where the authors modeled the result of a measurement process on a system initially in a state described by a Wigner function $F$, in terms of the (non-negative) phase space distribution obtained by smoothing
(convoluting) $F$ with another Wigner function $F_0$. For fixed $F_0$, this can be regarded as quantum dynamical map (a linear map that takes
Wigner functions to Wigner functions) as long as the convolution $F_0 \star F$ is a Wigner function for any Wigner function $F$. The authors
then proved that this will happen provided $F_0$ is point-wise non-negative. The procedure generates a large set of non-negative Wigner
functions. On the other hand, in refs. \cite{Varilly,Mukunda} necessary and sufficient conditions were derived for a Gaussian phase space
function to be a Wigner function. Gaussians are obviously important for their role in optics \cite{Sudarshan} and because they are {\it bona
fide} probability measures. Moreover, they provide kernels $F_0$ for quantum dynamical maps in the above sense. Strictly speaking,
Gracia-Bond\'{\i}a and V\'arilly \cite{Varilly} generalized the result of \cite{Simon} by proving that: (i) if $F_0$ is any point-wise
non-negative phase space function (Wigner function or not), then $F_0 \star F$ is a Wigner function for any Wigner function $F$, and (ii) if
$F_0$ and $F$ are Wigner functions, then $F_0 \star F$ is always non-negative (albeit possibly not a Wigner function). Using these results, they
constructed explicit examples that preclude the Hudson, Soto, Claverie theorem for mixed states; i.e. there are non-Gaussian positive Wigner
functions associated with mixed states.

Other authors considered whether certain criteria (namely the uncertainty principle) would be useful in assessing whether a phase space function is a Wigner function. In \cite{Manko,Gosson1} it was shown that the uncertainty principle does not determine the quantum state. Conditions for Gaussians in phase space to be Wigner distributions were expressed in terms of the so-called symplectic capacity of the associated Wigner ellipsoid \cite{Gosson2}. In \cite{Gosson3}, the multidimensional Hardy uncertainty principle was extended and expressed in the context of Wigner quasi-distributions.

The concept of the Narcowich-Wigner (NW) spectrum was introduced and explored in refs.\cite{Narcowich2}, \cite{Narcowich3}. Its main purpose was
to provide a criterion to classify certain phase space functions and a framework suitable to generate Wigner functions displaying specific
properties (namely being non-negative). With each square integrable phase space function $F$, we may associate its NW spectrum, denoted as
${\cal W} (F)$, which is a compact subset of $\bkR$. If this subset contains Planck's constant $\hbar$, then the phase space function $F$ is a
Wigner function and represents some quantum mechanical state. If, on the other hand, $0 \in {\cal W} (F)$, then $F$ is a point-wise non-negative
function, that is a classical probability measure. We see that, quite straightforwardly, the NW spectrum provides a criterion for questions i)
and iv) above.

On the other hand, the results of \cite{Simon}, \cite{Mukunda} were all reformulated and unified in the context of NW spectra in
ref.\cite{Narcowich3}. Various authors \cite{Narcowich2}, \cite{Werner} also speculated about the possibility of using the NW spectrum to
generate the entire subset of Wigner functions which are everywhere non-negative (i.e. contain $0$ in their NW spectra). While this has still
not been fully achieved, by resorting to the NW spectrum, Narcowich \cite{Narcowich2} proved that one can generate a large set of positive
Wigner functions of mixed states by convoluting suitable Wigner functions. This was basically a generalization of the previous results of
\cite{Simon} and \cite{Varilly}. Other potentially interesting applications of the NW spectrum are in the field of signal processing
\cite{Leonowicz} and in the study of the classical limit of quantum mechanics \cite{Werner2}.

An unresolved problem is that of using the NW spectrum to characterize pure and mixed states and thus to answer the questions ii) and iii)
above. In its general form this is still an open issue. For Gaussian states, however, the spectrum is of the form ${\cal W} (F) = \left[ -\eta_0
, \eta_0 \right]$ (eq.(20)) and it is known that $\eta_0=\hbar$ iff the state is pure and $\eta_0>\hbar$ iff it is mixed ($\eta_0 <\hbar$ iff
$F$ is not a Wigner function).

In this work we will prove an important result towards the solution of the above problem: we will complete the characterization of the NW
spectrum of pure states. In fact the purpose of this paper is to prove the following theorem.

\vspace{0.3 cm}
\noindent
{\bf Main Theorem
} {\it Let $F \in L^2 (\bkR^{2d} , d \xi)$ be the Wigner function of a pure state.

\vspace{0.3 cm}
\noindent
(i) If $F$ is a Gaussian, then ${\cal W} (F) = \left[- \hbar , \hbar \right]$.

\noindent
(ii) If $F$ is non-Gaussian, then ${\cal W} (F) = \left\{- \hbar, \hbar \right\}$.}

\vspace{0.3 cm}
\noindent
Notice that, as mentioned before, the result (i) in the Main Theorem is already known \cite{Narcowich2}. We include it here for completeness. 

We will also prove an auxilary result concerning the zeros of the so-called Husimi function \cite{Lee}. This result is stated in proposition 2.4.

The Main Theorem is of course consistent with the Hudson, Soto, Claverie theorem: Gaussian states are the only pure states that contain zero in their NW
spectrum. Unfortunately, the analogous characterization of the NW spectrum of mixed states remains an open issue. As shown in \cite{Werner},
mixed states can have a very intricate Wigner spectrum: intervals, sequences, or combinations of both.

This paper is organized as follows: Section 2 introduces the notation, the main definitions and reviews the fundamental properties of the NW
spectrum. In section 3 we prove our main results concerning the NW spectrum of pure states. We start by proving some auxiliary results for the Husimi function and review some properties of Bargmann transforms in section 3.1, and in section 3.2 we prove the Main Theorem. Finally, in section 4 we discuss some generalizations and possible applications of these results.

\section{The Narcowich-Wigner spectrum}

Let us first settle the preliminaries. We shall consider a $d$-dimensional system on a flat phase space $T^*M \simeq (\bkR^d)^* \times \bkR^d \simeq \bkR^{2d}$ with a global Darboux chart $\xi = (q,p) \in \bkR^{2d}$ in terms of which the symplectic form reads:
\begin{equation}
\Omega (\xi, \xi') = \xi^T {\bf J} \xi'= q \cdot p'- p \cdot q', \hspace{1 cm} {\bf J} = \left( \begin{array}{c c}
{\bf 0}_{d \times d} & {\bf I}_{d \times d}\\
- {\bf I}_{d \times d} & {\bf 0}_{d \times d}
\end{array}
\right)
\label{Eq1.1}
\end{equation}
Here the superscript "T" denotes matrix transposition. We shall use the standard norms in real, $||x||_{\bkR^n}= \sqrt{x_1^2 + \cdots + x_n^2}$, and in complex vector spaces $||z||_{\bkC^n}= \sqrt{|z_1|^2 + \cdots + |z_n|^2}$. To avoid a proliferation of subscripts, we shall just write $||x||$ or $||z||$, as it will be clear from the context the vector space we are referring to as well as its dimensionality. As usual $L^2 (\bkR^n, dx)$ denotes the space of square integrable complex-valued functions on $\bkR^n$ with respect to the Lebesgue measure. We may define in $L^2 (\bkR^n, dx)$ the inner product:
\begin{equation}
< \psi| \phi> = \int_{\bkR^n} \overline{\psi (x)} \phi (x) dx, \hspace{0.5 cm} \psi, \phi \in L^2 (\bkR^n,dx).
\label{Eq1.2}
\end{equation}
We shall also consider the vector space ${\cal T}$ of Hilbert-Schmidt operators acting on the Hilbert space ${\cal H}= L^2 (\bkR^d, dq)$ of our system, and which admit a kernel representation of the form:
\begin{equation}
\begin{array}{l l l l}
{\bf A}: & {\cal H} & \longrightarrow & {\cal H}\\
 & \psi (q) & \longmapsto & ({\bf A} \psi ) (q) = \int_{\bkR^d} A (q,q') \psi (q') d q',
\end{array}
\label{Eq1.3}
\end{equation}
with $A(q,q') \in L^2 (\bkR^{2d}, dqdq')$. The {\bf Weyl-Wigner transform} is the isomorphism \cite{Folland}:
\begin{equation}
\begin{array}
{l l l l}
W_{\eta}: & {\cal T}  & \longrightarrow & L^2 (\bkR^{2d} , d \xi)\\
& {\bf A} & \longmapsto & W_{\eta} ({\bf A}) (\xi) = \eta^d \int_{\bkR^d} e^{ - ip \cdot y} A \left(q + \frac{\eta y}{2}, q - \frac{\eta y}{2} \right) dy,
\end{array}
\label{Eq1.4}
\end{equation}
where $\eta$ is some real, positive constant. The Weyl-Wigner transform may be extended to other spaces, but this is all that will be necessary in this work. For a state $\psi (q) \in {\cal H}$, we define the corresponding density matrix to be the operator $ \bf{\rho}_{\psi} \in {\cal T}$:
\begin{equation}
({\bf \rho}_{\psi} \phi) (q) = \int_{\bkR^d} r_{\psi} (q,q') \phi (q') d q',
\label{Eq1.5}
\end{equation}
with $r_{\psi} (q,q') = \psi (q) \overline{\psi (q')}$. The $\eta$-Wigner function associated with $\psi$ is \cite{Wigner}:
\begin{equation}
W_{\eta} (\psi, \psi) (\xi) = \frac{1}{(2 \pi \eta)^d} W_{\eta} (\rho_{\psi}) (\xi) = \frac{1}{(2 \pi \eta)^d} \int_{\bkR^d} e^{-2ip \cdot y/ \eta} \overline{\psi (q-y)} \psi (q+y) dy.
\label{Eq1.6}
\end{equation}
If $\eta$ is equal to Planck's constant $\hbar$, we shall denote $W_{\hbar} (\psi, \psi) (\xi)$ simply by {\bf Wigner function} or {\bf Wigner quasi-distribution} in accordance with the literature. Moreover, it is straightforward to prove the following useful identity, also known as the Moyal identity \cite{Moyal}:
\begin{equation}
\int_{\bkR^{2d}} \overline{W_{\eta}(\psi, \psi) (\xi)} W_{\eta} (\phi, \phi)(\xi) d \xi = \frac{1}{(2\pi \eta)^d  } |<\psi| \phi>|^2,
\label{Eq1.7}
\end{equation}
where $W_{\eta} (\psi, \psi), W_{\eta} (\phi, \phi)$ are $\eta$-Wigner functions associated with states $\psi, \phi \in L^2 (\bkR^d, dq)$.

We may consider statistical mixtures of pure states:
\begin{equation}
{\bf \rho} = \sum_{\alpha} p_{\alpha} {\bf \rho}_{\psi_{\alpha}} , \hspace{0.5 cm} p_{\alpha} \ge 0, ~ \sum_{\alpha} p_{\alpha} =1,
\label{Eq1.8}
\end{equation}
which yield, via the Weyl-Wigner transform, the mixed state $\eta$-Wigner functions:
\begin{equation}
W_{\eta} ({\bf \rho}) (\xi) = \sum_{\alpha}  p_{\alpha} W_{\eta} (\psi_{\alpha}, \psi_{\alpha}) (\xi) .
\label{Eq1.9}
\end{equation}
The Wigner functions thus represent the states in the quantum phase space. An important issue is the identification of necessary and sufficient conditions for a phase space function to be a $\eta$-Wigner function \cite{Dias}. A well-known necessary condition is stated in the following proposition \cite{Dias}.

\vspace{0.3 cm}
\noindent
{\bf Proposition 1.1} {\it If $F^{\eta}$ is a $\eta$-Wigner function, then:
\begin{equation}
\int_{\bkR^{2d}} |F^{\eta} (\xi)|^2 d \xi \le \frac{1}{(2 \pi \eta)^d}.
\label{Eq1.10}
\end{equation}
In addition, the equality holds if and only if $F^{\eta}$ is associated with a pure state.}

\vspace{0.3 cm}
\noindent
The integral on the left-hand side of the previous inequality is called the purity of the system.

The concept of NW spectrum was introduced in ref.\cite{Narcowich2}. It appears naturally if one wants to formulate necessary and sufficient conditions (the KLM conditions) for a phase space function to be a Wigner function or, more generally, a $\eta$-Wigner function. In addition, we are able to (partially) characterize different types of Wigner functions in terms of their NW spectra, namely the non-negative Wigner functions, the Gaussians, and pure or mixed states \cite{Narcowich3}. For this purpose it is useful to define the {\bf symplectic Fourier transform}. Let $G(\xi) \in L^2( \bkR^{2d} , d \xi)$. We define its symplectic Fourier transform according to:
\begin{equation}
{\cal F}: L^2( \bkR^{2d}) \to L^2( \bkR^{2d}): \hspace{0.5 cm} G(\xi) \mapsto \hat G (a) \equiv ({\cal F} G) (a) = \int_{\bkR^{2d}} ~ G(\xi) \exp \left( i \Omega (\xi, a) \right) d \xi.
\label{Eq1.11}
\end{equation}
The inverse transform is given by the formula:
\begin{equation}
G (\xi) \equiv ({\cal F}^{-1}  \hat G) (\xi) = \frac{1}{(2 \pi)^{2d}} \int_{\bkR^{2d}} ~ \hat G(a) \exp \left( - i \Omega (\xi, a) \right) d a.
\label{Eq1.12}
\end{equation}

\vspace{0.3 cm}
\noindent
{\bf Definition 1.2} {\it The symplectic Fourier transform $\hat F(a)$ is said to be of the {\bf $\alpha$-positive type} $(\alpha \in \bkR)$ if the $m \times m$ matrix with entries
\begin{equation}
M_{jk}= \hat F (a_j -a_k) \exp \left( \frac{i \alpha}{2} \Omega (a_j , a_k ) \right)
\label{Eq1.13}
\end{equation}
is hermitian and non-negative for any positive integer $m$ and any set of $m$ points $a_1, \cdots, a_m$ in the dual of the phase space. By abuse of language, we shall often say that a phase space function $F(\xi)$ is of the $\alpha$-positive type, by which we mean that its symplectic Fourier transform $\hat F (a)$ is of the $\alpha$-positive type. The {\bf Narcowich-Wigner spectrum} of a phase space function $F(\xi) \in L^2 ( \bkR^{2d} , d \xi)$ is the set:}
\begin{equation}
{\cal W} (F) = \left\{ \alpha \in \bkR \left| \hat F (a) {\mbox{ is of the $\alpha$-positive type}} \right. \right\}.
\label{Eq1.14}
\end{equation}

\vspace{0.3 cm}
\noindent
With this definition we may now state the KLM (Kastler, Loupias, Miracle-Sole \cite{Kastler}, \cite{Loupias}, \cite{Narcowich1}) conditions.

\vspace{0.3 cm}
\noindent
{\bf Theorem 1.3} {\it The phase space function $F(\xi)$ is a $\eta$-Wigner function ($\eta >0$), iff its symplectic Fourier transform $\hat F (a)$ satisfies the KLM conditions:
\begin{eqnarray}
(i) & \hat F (0)  & =1 \label{Eq1.15} \\
(ii) & \hat F (a) & {\mbox{is continuous and $\eta \in {\cal W} (F)$.}} \label{Eq1.16}
\end{eqnarray}}

\vspace{0.3 cm}
\noindent
This theorem is a twisted generalization of Bochner's theorem, which states that non-negative functions (i.e. classical probability densities) are those of $0$-positive type. To proceed we introduce
the convolution of $F,G \in L^2 (\bkR^{2d}, d \xi)$ which is defined by:
\begin{equation}
(F * G) (\xi) = \int_{\bkR^{2d}} ~ F(\xi- \xi') G (\xi') d \xi',
\label{Eq1.17}
\end{equation}
and is continuous, bounded and vanishes  for $|| \xi|| \to \infty$. Moreover, the symplectic Fourier transform of the convolution is
tantamount to point-wise multiplication:
\begin{equation}
({\cal F} (F*G)) (a) = ({\cal F} (F)) (a) \cdot ({\cal F} (G)) (a).
\label{Eq1.18}
\end{equation}
Let us now recapitulate some of the properties of the NW spectrum.

\vspace{0.3 cm}
\noindent
{\bf Proposition 1.4} {\it Let $F,G \in L^2 (\bkR^{2d}, d \xi)$.

\vspace{0.3 cm}
\noindent
(i) If $f$ is continuous, then ${\cal W} (F)$ is bounded and closed

\noindent
(ii) $\alpha \in {\cal W} (F)\Leftrightarrow (- \alpha) \in {\cal W} (F)$

\noindent
(iii) $\left\{\alpha + \alpha'| ~ \alpha \in {\cal W} (F),~ \alpha' \in {\cal W} (G) \right\}\subseteq {\cal W} (F*G)$, where $(F*G)$ is the convolution in (\ref{Eq1.17}).}

\vspace{0.3 cm}
\noindent
The proof of these properties can be found in \cite{Werner}. Counter-examples to the reciprocal of property (iii) are also given in ref.\cite{Werner}.
In view of property (ii) of proposition 1.4, we may assume without loss of generality that $\alpha \ge 0$. The results of \cite{Varilly}, \cite{Mukunda} for Gaussian states can be expressed in terms of the NW spectrum as follows (see \cite{Narcowich2} for a proof):

\vspace{0.3 cm}
\noindent
{\bf Lemma 1.5} {\it Let $F$ be a Gaussian
\begin{equation}
F(\xi) = \frac{\sqrt{\det {\bf A}}}{\pi^d} \exp \left[ - (\xi - \xi_0)^T {\bf A} (\xi - \xi_0) \right],
\label{Eq1.19}
\end{equation}
where ${\bf A}$ is a real, symmetric, positive definite $2d \times 2d$ matrix and $\xi_0 \in \bkR^{2d}$. Then the NW spectrum of $F$ is of the form:
\begin{equation}
{\cal W} (F) = \left[ - \eta_0 , \eta_0 \right],
\label{Eq1.20}
\end{equation}
where $\eta_0 = max({\cal W} (F))$. That this maximum exists is guaranteed by property (i) of proposition 1.4.}

\vspace{0.3 cm}
\noindent
The following lemma states the well known results of \cite{Simon}, \cite{Varilly} mentioned in the introduction. We include it here to shown the usefulness of the NW spectra. Indeed, using the previous results, the proof is trivial, as shown in \cite{Narcowich3}.

\vspace{0.3 cm}
\noindent
{\bf Lemma 1.6} {\it Let $F$, $F_0$ be Wigner functions.

\vspace{0.3 cm}
\noindent
(i) The convolution $F_0 * F$ is point-wise non-negative.

\vspace{0.3 cm}
\noindent
(ii) If $F_0$ is such that $0 \in {\cal W} (F_0)$ or $2 \hbar \in {\cal W} (F_0)$, then the convolution $F_0 * F$ is a Wigner function.
}

\vspace{0.3 cm}
\noindent
{\bf Proof} If $F,F_0$ are Wigner functions, then $\left\{- \hbar, \hbar \right\} \subset {\cal W} (F_0) \cap {\cal W} (F)$ and thus, from proposition 1.4 (iii) $0 \in {\cal W} (F_0 \star F)$, which means that $F_0 \star F$ is point-wise non-negative.
On the other hand, if $F_0$ is such that $\left\{0, \pm \hbar \right\} \subset {\cal W} (F_0)$ or $\left\{\pm \hbar, \pm 2 \hbar \right\} \subset {\cal W} (F_0)$, then from proposition 1.4 (iii), $\left\{- \hbar, 0, \hbar \right\} \subset {\cal W} (F_0 * F)$, and thus $F_0 \star F$ is a Wigner function and point-wise non-negative.$_{\Box}$

\vspace{0.3 cm}
\noindent
This is a powerful way of generating positive Wigner functions, notwithstanding the fact that the method does not exhaust the entire set of positive Wigner functions.

\section{Main results}

\subsection{Husimi function and Bargmann transform}

We start by deriving some results which will be useful to prove the Main Theorem. The following restricted version of Hadamard's theorem for functions of several complex variables was proved in \cite{Soto}:

\vspace{0.3 cm}
\noindent
{\bf Theorem 2.1 (Hadamard, Soto, Claverie)} {\it If $F(z)$ is an entire function on $\bkC^n$, with order of growth $\rho$ and without zeroes, we have:
\begin{equation}
F(z) = \exp \left(P(z) \right)
\label{Eq2.1}
\end{equation}
where $P(z)$ is a polynomial of degree $r \le \rho$.}

\vspace{0.3 cm}
\noindent
We recall that the order of growth $\rho$ of $F(z)$ is given by:
\begin{equation}
\rho = \lim_{R \to \infty} \frac{\log \left( \log M (R) \right)}{\log R}, \hspace{1 cm} M(R) = \begin{array}{c l}
{\mbox{sup}} & |F(z)|\\
||z||=R &
\end{array}
\label{Eq2.2}
\end{equation}
Let us now consider the coherent state
\begin{equation}
\psi_z (q) = (\pi \hbar)^{-\frac{d}{4}} \exp \left( - \frac{||q||^2}{2 \hbar} + z \cdot q - \frac{\hbar}{2} ||Re z||^2 \right), \hspace{0.5 cm} z \in \bkC^d.
\label{Eq2.3}
\end{equation}
Moreover, let $W_{\hbar} (\psi_z, \psi_z)$ and $W_{\hbar} (\psi, \psi)$ denote the Wigner functions associated with $\psi_z$ and $\psi \in L^2 (\bkR^d, dq)$, respectively.

\vspace{0.3 cm}
\noindent
{\bf Definition 2.2} Let $\psi \in L^2 (\bkR^d, dq)$. The convolution of the form
\begin{equation}
Q(\xi) = (W_{\hbar} (\psi_z, \psi_z)* W_{\hbar} (\psi, \psi)) (\xi)
\label{Eq2.4}
\end{equation}
is called a {\bf Husimi function}. The {\bf Bargmann transform} of $\psi$ is defined by \cite{Folland}, \cite{Bargmann}:
\begin{equation}
F(z) \equiv < \psi| \psi_z> e^{\frac{\hbar}{2} ||Rez||^2} = \frac{1}{(\pi \hbar)^d} \int_{\bkR^d} \exp \left(-\frac{||q||^2}{2 \hbar} + z \cdot q \right) \overline{\psi (q)} dq, \hspace{0.5 cm} z \in \bkC^d.
\label{Eq2.5}
\end{equation}

\vspace{0.3 cm}
\noindent
The following lemma provides an important characterization of the Bargmann transform:

\vspace{0.3 cm}
\noindent
{\bf Lemma 2.3} {\it Let $\psi \in L^2 (\bkR^d, dq)$ and let $F(z)$ be its Bargmann transform. Then $F(z)$ is an entire function on $\bkC^d$. Moreover, the order of growth $\rho$ of $F(z)$ is at most 2.}

\vspace{0.3 cm}
\noindent
{\bf Proof} Since the integral in (\ref{Eq2.5}) converges uniformly in any compact subset of $\bkC^d$, we conclude that $F(z)$ is an entire function on $\bkC^d$. On the other hand, since $\psi, \psi_z$ are both normalized, we have from the Cauchy-Schwarz inequality:
\begin{equation}
|F(z) |= e^{\frac{\hbar}{2} ||Rez||^2} | < \psi| \psi_z>| \le e^{\frac{\hbar}{2} ||Rez||^2}.
\label{Eq2.6}
\end{equation}
Consequently, $M(R) \le e^{\frac{\hbar}{2} R^2}$ and thus $\rho \le 2$.$_{\Box}$

\vspace{0.3 cm}
\noindent
The following result will be useful to prove the Main Theorem.

\vspace{0.3 cm}
\noindent
{\bf Proposition 2.4} {\it The Husimi function $Q(\xi)$ has no zeroes iff the state $\psi \in L^2 (\bkR^d, dq)$ is a Gaussian.}

\vspace{0.3 cm}
\noindent
{\bf Proof}
If we compute $W_{\hbar} (\psi_z, \psi_z)$ explicitly (cf.(\ref{Eq1.6},\ref{Eq2.3})), we obtain:
\begin{equation}
W_{\hbar} (\psi_z, \psi_z)(\xi) = \frac{1}{(\pi \hbar)^d} \exp \left(- \frac{1}{\hbar} ||\xi - \zeta||^2 \right)
\label{Eq2.7}
\end{equation}
where
\begin{equation}
\zeta = (x,k) = (\hbar {\mbox { Re}} (z), \hbar {\mbox { Im}} (z)).
\label{Eq2.8}
\end{equation}
The convolution (\ref{Eq2.4}) thus reads:
\begin{equation}
Q(\xi) = \frac{1}{(\pi \hbar)^d} \int_{\bkR^{2d}} \exp \left( - \frac{1}{\hbar} ||\xi - \zeta - \xi'||^2 \right) W_{\hbar} (\psi, \psi) (\xi') d \xi'.
\label{Eq2.9}
\end{equation}
We conclude that if $\psi$ is a Gaussian, then $W_{\hbar} (\psi, \psi)$ is a Gaussian and so is $Q$. Then, evidently, $Q$ has no zeroes.

Conversely, let us assume that $\psi$ is not a Gaussian, but that it has no zeroes. This means that the Bargmann transform $F(z)$ also has no zeroes. Indeed, we have:
\begin{equation}
\begin{array}{c}
|F(z)|^2 = e^{\hbar ||Rez||^2} | < \psi| \psi_z>|^2= (2 \pi \hbar)^d e^{\hbar ||Rez||^2} \int_{\bkR^{2d}} \overline{W_{\hbar} (\psi_z, \psi_z) (\xi)} W_{\hbar} (\psi, \psi) (\xi) d \xi = \\
\\
= 2^d e^{\hbar ||Rez||^2} \int_{\bkR^{2d}} \exp \left(- \frac{1}{\hbar} ||\zeta - \xi'||^2 \right) W_{\hbar} (\psi, \psi) (\xi') d \xi' = (2 \pi \hbar)^d  e^{\hbar ||Rez||^2} Q (2 \zeta),
\label{Eq2.10}
\end{array}
\end{equation}
where we used (\ref{Eq1.7}). This means that $Q$ has zeroes iff $F(z)$ has zeroes. From the Hadamard, Soto Claverie theorem (Theorem 2.1), since $F(z)$ is an entire function on $\bkC^d$, with order of growth $\rho \le2$ and, by hypothesis, without zeroes, then it must be a Gaussian. We then have, in particular, that:
\begin{equation}
F(i y) = \frac{1}{(\pi \hbar)^{\frac{d}{4}}} \int_{\bkR^d} \exp \left( -\frac{ ||q||^2}{2 \hbar} + i y \cdot q \right) \overline{\psi (q)} dq , \hspace{0.5 cm} y \in \bkR^d,
\label{Eq2.11}
\end{equation}
is a Gaussian. As $F(iy)$ is the Fourier transform of $e^{- \frac{||q||^2}{2 \hbar}} \overline{\psi (q)}$, this is only possible if $e^{-\frac{||q||^2}{2 \hbar}} \overline{\psi (q)}$ is also a Gaussian. Hence $\psi (q)$ is a Gaussian, which contradicts our assumption.$_{\Box}$

\subsection{Proof of the Main Theorem}

\vspace{0.3 cm}
\noindent
The following lemma restricts the NW spectra of pure states.

\vspace{0.3 cm}
\noindent
{\bf Lemma 2.5} {\it Let $F$ be the Wigner function of a pure state. Then:}
\begin{equation}
{\cal W} (F) \subseteq \left[ - \hbar, \hbar \right]
\label{Eq2.12}
\end{equation}

\vspace{0.3 cm}
\noindent
{\bf Proof} From proposition 1.1, we know that, since $F$ is a Wigner function the equality in (\ref{Eq1.10}) holds for $\eta= \hbar$. Moreover, if $\eta >0$ is some other element in ${\cal W} (F)$, then again from (\ref{Eq1.10}), we know that:
\begin{equation}
\frac{1}{(2 \pi \hbar)^d}= \int_{\bkR^{2d}} |F(\xi)|^2 d \xi \le \frac{1}{(2 \pi \eta)^d}.
\label{Eq2.13}
\end{equation}
It follows that $0< \eta \le \hbar$. The lemma is then an immediate consequence of property (ii) of proposition 1.4.$_{\Box}$

\vspace{0.3 cm}
\noindent
The main theorem of this work refines the previous lemma.

\vspace{0.3 cm}
\noindent
{\bf Proof of the Main Theorem}

\noindent
(i) This is well-known. We prove it here for completeness. Let $F$ be a Gaussian (\ref{Eq1.19}). From lemma 1.5, we know that for $F$ to be a Wigner function (pure or mixed), then the inclusion must hold:
\begin{equation}
\left[- \hbar, \hbar \right] \subseteq \left[- \eta_0 , \eta_0 \right] = {\cal W} (F).
\label{Eq2.14}
\end{equation}
But from lemma 2.5, if $F$ is associated with a pure state, then the inverse inclusion must also hold. We conclude that:
\begin{equation}
{\cal W} (F) = \left[- \hbar, \hbar \right]
\label{Eq2.15}
\end{equation}

\vspace{0.3 cm}
\noindent
(ii) Let us now assume that $F(\xi)$ is the Wigner function of a pure state, but that it is not a Gaussian. We already know from lemma 2.5, that its NW spectrum is contained in $\left[- \hbar, \hbar \right]$. Let us then assume that there exists $\eta \in \left. \right] 0, \hbar \left[ \right. \cap {\cal W} (F)$. Let us consider the family of Gaussians:
\begin{equation}
G_{\alpha} (\xi) = \frac{1}{(\pi \alpha)^d} \exp \left(-\frac{1}{\alpha} || \xi  ||^2 \right), \hspace{0.5 cm} \alpha >0
\label{Eq2.16}
\end{equation}
The NW spectrum is ${\cal W} (G_{\alpha} ) = \left[- \alpha, \alpha \right]$ \cite{Narcowich2}. The convolution of two such Gaussians is again a Gaussian in this family. In fact, the convolution verifies the semi-group law:
\begin{equation}
(G_{\alpha} * G_{\beta} ) (\xi) = G_{\alpha + \beta} (\xi), \hspace{0.5 cm} \alpha, \beta >0
\label{Eq2.17}
\end{equation}
Let us then define the function:
\begin{equation}
Q(\xi) \equiv (G_{\hbar - \eta} * G_{\eta} * F) (\xi) = (G_{\hbar} * F) (\xi) = (G_{\hbar - \eta} * Q_0 ) (\xi),
\label{Eq2.18}
\end{equation}
where $Q_0 (\xi) = (G_{\eta} * f) (\xi)$.
The Wigner spectrum of $G_{\eta}$ is ${\cal W} (G_{\eta}) = \left[- \eta, \eta \right]$, whereas that of $F$ contains the set $\left\{- \hbar, - \eta, \eta , \hbar \right\}$ by assumption. From property (iii) of proposition 1.4, we conclude that $\left\{- \hbar, 0, \hbar \right\} \subset {\cal W} (Q_0)$. This means that $Q_0$ is both a Wigner function and, according to Bochner's theorem, everywhere non-negative. Since $Q(\xi) = (G_{\hbar} * F) (\xi)$ is the convolution of a Gaussian of width $\hbar$ with a Wigner function, it is a Husimi function. Moreover, since $F$ is not a Gaussian, then $Q(\xi)$ has at least one zero $\xi_0$ (proposition 2.4):
\begin{equation}
Q(\xi_0)=0
\label{Eq2.19}
\end{equation}
From (\ref{Eq2.18}), we then have:
\begin{equation}
0 = \int_{\bkR^{2d}} \exp \left( - \frac{||\xi_0 - \xi||^2}{\hbar - \eta} \right) Q_0 (\xi) d \xi.
\label{Eq2.20}
\end{equation}
This is only possible, if $Q_0 (\xi)$ has a negative part. However, that is contradictory with the fact that $Q_0$ is of $0$-positive type. This proves that there can be no $\eta \in \left. \right] 0 ,\hbar \left[ \right. \cap {\cal W} (F)$. Altogether, this means that the NW spectrum of a non-Gaussian pure state is contained in $\left\{- \hbar, 0, \hbar \right\}$. However, since non-Gaussian pure states cannot have non-negative Wigner functions (by the Hudson, Soto, Claverie Theorem), we conclude that $0 \notin {\cal W} (F)$ and thus ${\cal W} (F)=\left\{- \hbar,  \hbar \right\}$.$_{\Box}$

\section{Concluding Remarks}

\vspace{0.3 cm}
\noindent
A few remarks are now in order:

\vspace{0.3 cm}
\noindent
$\bullet$ The Main Theorem admits a trivial generalization. We can easily replace $\hbar$ by any $\eta>0$. And thus, if $\psi$ is a Gaussian, then the $\eta$-Wigner function $W_{\eta} (\psi, \psi) (\xi)$ has Wigner spectrum $\left[- \eta, \eta \right]$, otherwise its Wigner spectrum is $\left\{- \eta, \eta \right\}$.

\vspace{0.3 cm} \noindent $\bullet$ The converse result of the Main Theorem remains as an open question. That is: suppose $F \in L^2 (\bkR^{2d} , d
\xi)$ is such that $\hat F (a)$ is continuous and has NW spectrum ${\cal W} (F) = \left\{- \hbar, \hbar \right\}$. Does that mean that $F$ is
the Wigner function associated with a non-Gaussian pure state? Or could it be a mixed state? If we could prove this result then pure and mixed
states could be distinguished exclusively in terms of their NW spectra.

\vspace{0.3 cm}
\noindent
$\bullet$ In \cite{Narcowich2} Narcowich stated the following conjecture:

\vspace{0.3 cm}
\noindent
{\it The convolution of a fixed Wigner function $F^0_{\hbar}$ with any other Wigner function $F_{\hbar}$ is again a Wigner function iff the NW spectrum of $F^0_{\hbar}$ contains $0$ or $\pm 2 \hbar$.}

\vspace{0.3 cm}
\noindent
As we mentioned before, the sufficiency of the conditions is trivial to prove (see lemma 1.6). Narcowich argued that the necessity is proved, if we assume that (i) ${\cal W} (F * G) = {\cal W} (F)+ {\cal W} (G) \equiv \left\{\alpha + \alpha' | \alpha \in {\cal W} (F), ~ \alpha' \in {\cal W} (G) \right\}$ for any $F,G \in L^2 (\bkR^{2d}, d \xi)$; and (ii) there exists a family of Wigner functions $F^{\epsilon}_{\hbar} (\xi)$ defined for all sufficiently small $\epsilon >0$ and satisfying ${\cal W} (F^{\epsilon}_{\hbar}) \subseteq \left[- \hbar - \epsilon, -\hbar + \epsilon \right] \cup \left[\hbar - \epsilon, \hbar + \epsilon \right]$. Let us reproduce here his argument for completeness. If $F^0_{\hbar}$ is a Wigner function and $F^0_{\hbar} * F_{\hbar}$ is yet another Wigner function for any Wigner function $F_{\hbar}$, then $\hbar \in {\cal W} (F^0_{\hbar} * F_{\hbar})$. From hypothesis (i), it follows that $\hbar \in {\cal W} (F^0_{\hbar}) + {\cal W} (F_{\hbar}^{\epsilon})$. If we take $\epsilon \searrow 0$, we get $\hbar \in {\cal W} (F^0_{\hbar}) + \left\{- \hbar, \hbar \right\}$, which is possible if and only if either $0$ or $\pm 2 \hbar$ belong to ${\cal W} (F^0_{\hbar})$. The problem with this argument is that, as mentioned before, hypothesis (i) is not true in general \cite{Werner}. However, that does not mean that the conjectured theorem is wrong. To the best of our knowledge no counter-example has ever been found. The problem might be that the hypotheses suggested may be too restrictive. Let us just assume that for the fixed Wigner function $F^0_{\hbar}$, we can always find a non-Gaussian pure state $\psi$ such that ${\cal W} (F^0_{\hbar} * W_{\hbar} (\psi, \psi)) = {\cal W} (F^0_{\hbar})+ {\cal W} (W_{\hbar} (\psi, \psi))$. Notice that we do not require this to be true for all Wigner functions. We only need this to be true for one state $W_{\hbar} (\psi, \psi)$. We then have, from the Main Theorem: $\hbar \in {\cal W} (F^0_{\hbar} * W_{\hbar} (\psi, \psi) ) = {\cal W} (F^0_{\hbar} ) + \left\{- \hbar, + \hbar \right\}$. And the necessity would thus be proved. The main result of this paper shows that at least hypothesis (ii) is easily met.

\vspace{0.3 cm}
\noindent
$\bullet$ Our lemma 2.5 has an important consequence in the context of open (dissipative quantum systems). In this framework one usually regards the purity (\ref{Eq1.10}) of the system interacting with an external environment \cite{Zeh} as an indication of dissipative effects. In fact it tends to decrease. In most cases if $F_{\hbar} (\xi)$ is the Wigner function of the system at the initial time, then the Wigner function will usually take the form $G_t(\xi) * F_{\hbar} (\xi)$ at a later time $t$ (up to composition with a linear symplectic transformation, which leaves the NW spectrum unchanged \cite{Narcowich2}). Here $G_t (\xi)$ is the Green function of the master equation. If the Narcowich conjecture mentioned above is correct, then $G_t$ should be such that $\pm 2 \hbar \in {\cal W} (G_t)$ or, alternatively that $0 \in {\cal W} (G_t)$. Moreover, if $F_{\hbar}$ is associated with a pure state ($F_{\hbar}= W_{\hbar} (\psi, \psi)$), then from lemma 2.5, we infer that $G_t*F_{\hbar}$ must be associated with a mixed state, whenever there exists $\eta \ne 0$ in ${\cal W} (G_t)$.

\vspace{1 cm}

\begin{center}

{\large{{\bf Acknowledgments}}}

\end{center}

\vspace{0.3 cm}
\noindent
This work was partially supported by the grants PTDC/MAT/69635/2006 and POCTI/0208/2003 of the Portuguese Science Foundation.


\begin{thebibliography}{99}
\bibitem{Gosson3} M. de Gosson, F. Luef, "The multi-dimensional Hardy uncertainty principle and its interpretation in terms of the Wigner distribution; relation with the notion of symplectic capacity", math-ph/08030910.
\end{thebibliography}
\end{document}